\documentclass{PoS}

\def\la{\mathrel{\hbox{\rlap{\hbox{\lower4pt\hbox{$\sim$}}}\hbox{$<$}}}}
\def\ga{\mathrel{\hbox{\rlap{\hbox{\lower4pt\hbox{$\sim$}}}\hbox{$>$}}}}

\def\arcmin{\hbox{$^\prime$}}

\def\fm{\hbox{$.\!\!^{\rm m}$}}

\def\fdg{\hbox{$.\!\!^\circ$}}

\def\micron{\hbox{$\mu$m}}

\newcommand{\MSUN}{${\rm M}_\odot$}

\newcommand{\LSUN}{${\rm L}_\odot$}

\newcommand{\kms}{{\,km\,s$^{-1}$}}

\newcommand{\HI}{\mbox{\normalsize H\thinspace\footnotesize I}}

\def\apj    {{ApJ }}

\title{Properties of the extremely HI-massive galaxy HIZOA\,J0836-43}

\ShortTitle{Properties of the extremely HI-massive galaxy HIZOA\,J0836-43}

\author{\speaker{Ren\'ee C.Kraan-Korteweg}\\
       Astronomy Department, University of Cape Town\\
       E-mail: \email{kraan@ast.uct.ac.za}}

\author{Michelle E. Cluver\\
        IPAC, California Institute of Technology\\
        E-mail: \email{mcluver@ipac.caltech.edu}}

\author{Tom H. Jarrett\\
        IPAC, California Institute of Technology\\
        E-mail: \email{jarrett@ipac.caltech.edu}}

\author{Patrick A. Woudt\\
       Astronomy Department, University of Cape Town\\
        E-mail: \email{pwoudt@ast.uct.ac.za}}

      \abstract{Little is known about the properties of extremely
        massive HI-galaxies. They are extremely scarce and are --
        according to hierarchical structure formation -- only forming
        now ($z < 1$). The forthcoming deep HI SKA Pathfinders surveys
        will uncover many more of them. This will lead to a better
        understanding of their evolution and frequency, and the shape
        of the bright end of the HI mass function.

        The recently discovered galaxy HIZOA\,J0836-43 is one of the
        most HI-rich galaxies ($M_{\rm{HI}}$=7.5$\times10^{10}$\MSUN)
        -- and the nearest of its kind. As such it is an ideal local
        probe of these elusive galaxies. Results from a detailed
        investigation in the near- ({\sl IRSF}) and far-infrared ({\sl
          Spitzer}) of this local HI-massive galaxy are
        presented. Unlike other giant HI galaxies, it is not of low
        surface brightness. The galaxy is found to be a luminous
        starbursting galaxy at an unexpected early stage of stellar
        mass building, more typical of star-forming galaxies at higher
        redshift ($z\sim 0.7$). With regard to its environment, hence
        possible clues to its formation, the near infrared imaging
        survey finds HIZOA\,J0836-43 to lie in a region underdense in
        $L{^*}$ galaxies -- consistent with the observation that
        HI-massive galaxies are preferentially found in low density
        regions -- in the presence, however, of an uncommonly large
        number of low stellar mass galaxies.}

\FullConference{Panoramic Radio Astronomy: Wide-field 1-2 GHz research on galaxy evolution\\
		 June 2-5 2009\\
		 Groningen, the Netherlands}
\begin{document}
\section{Introduction}
The number of galaxies with HI-masses above a characteristic value,
$M^*_{\rm{HI}} = 6\cdot10^9$\MSUN\  declines exponentially (Zwaan et
al. 2005). The most massive galaxies are very scarce, their number density
is ill-constrained, and little is known about their properties.

Hierarchical structure formation predicts these galaxies to be forming
only recently ($z < 1$; Mo et al. 1998) although models overpredict
their current population ($z = 0 - 1$; Renzini 2007). Local massive
and gas-rich galaxies are thus expected to be in an early stage of
formation. But the very few known ones - like Malin 1 (Impey \& Bothun
1989) - are rather quiescent, providing few clues about their past or
future evolution.

A detailed investigation of the recently discovered, fairly nearby
($v_{hel}$=10689\kms) galaxy HIZOA\,J0843-36 (Donley et al. 2006), one
of the most massive HI-galaxies known to date, will therefore provide
a valuable local probe to be put into context with the forthcoming
higher redshift pathfinder surveys.

HIZOA\,J0843-36 was discovered with the Parkes radio telescope in the
deep systematic HI survey of the Zone of Avoidance (ZOA; e.g.
Kraan-Korteweg et al. 2005).  Follow-up ATCA and AAT radio
observations confirmed it to be amongst the most massive spiral
galaxies (Donley et al. 2006). It contains 7.5$\times10^{10}$\MSUN\ of
\HI\ gas, has a total dynamical mass of 1.4$\times10^{12}$\MSUN\ and a
20-cm derived star formation rate of $\sim$35\MSUN/yr. It also has a
prominent bulge in the near-infrared (bulge-to-disk ratio in the
$K_{s}$ band of $\sim 0.80$) which lies central to its enormous,
rapidly rotating HI disk. It is amongst the most HI-massive spiral
galaxies known to date and, at a distance of 148 Mpc, also the
nearest. Contrary to other HI-rich galaxies, like Malin 1, it is,
however, not a giant low surface brightness galaxy.

This leads to the question: What is happening in this unusual galaxy?
Is the galaxy undergoing a starburst, or is it an AGN? Can its
environment provide clues about its origin/formation? How did it grow
to its current size?  Was it through merger/accretion processes?

\section{Near and Mid-Infrared observations of HIZOA\,J0836-43}

A detailed investigation of HIZOA\,J0843-3 is rendered difficult due
to its location behind the Vela supernova remnant in the ZOA. Lying behind a
dust layer of $A_B = 10\fm0$ it is optically invisible. To minimize
the effects of foreground extinction, we launched an observational
campaign in the infrared of the galaxy and its local environment.  The
program comprises imaging of an area of $28\arcmin \times 28\arcmin$,
centered on the galaxy, obtained by using {\sl Spitzer} with IRAC (3.6,
4.5, 5.8, 8.0$\micron$) and MIPS (24 .0, 70, 160$\micron$), as well as
IRS high and low resolution spectroscopy of the galaxy itself. We also
performed a deep imaging survey ($J H K_s$) of $1\fdg5 \times 1\fdg5$
with the Infrared Survey Facility (IRSF) at the 1.4 m Japanese
telescope in Sutherland (SAAO), see Cluver (2009) for details.

\subsection{Infrared properties of HIZOA\,J0836-43}
The near-infrared light profiles of the galaxy appear typical for an
early spiral or lenticular with its prominent bulge, while the radio
observations suggest recent, active star formation. The new data find
a NIR -- MIR Spectral Energy Distribution that is consistent with an
Sc morphology (from GRASIL modeling). However, the total IR luminosity
of $L_{\rm{TIR}}$=$1.2\times{10}^{11}$\LSUN\ and star-formation rate
in the far infrared (FIR) of $\sim 21$\MSUN/yr is much higher than
that of typical disk spirals and is more consistent with the
properties of local Luminous Infrared Galaxies (LIRGs). The galaxy
also has substantial cold molecular gas
($M_{\rm{H_{2}}}$=$1.3\times{10}^{10}$\MSUN) as estimated from the
FIR, but little warm dust. The galaxy possesses a prominent bulge of
evolved stars and a total stellar mass of 4.4$\times10^{10}$\MSUN\
(Cluver et al. 2009).

In summary, we have a starbursting galaxy (the spectrum shows no signs
of AGN activity) with a prominent bulge comprised of evolved
stars. But its specific star formation rate ($\sim$0.5 $\rm{Gyr}^{-1}$
) implies an early stage of stellar mass building. With its huge
reservoir reservoir of gas it could double its stellar mass in about 2
Gyr.  Based on the stellar surface brightness profile and emission
from tracers of star formation, the galaxy seems to be undergoing
inside-out star formation similar to processes observed at higher
redshift ($z\sim 0.7$) where gas fractions of disks were higher
compared to local galaxies (e.g. Bell et al. 2005,
P\'{e}rez-Gonz\'{a}lez et al. 2005).  Despite its extreme \HI\ mass,
both the near- and mid-infrared luminosity appear consistent with
expectations from \HI/luminosity relations in these bands. This
suggests relatively ``normal'' evolution in HIZOA J0836-43, despite
lying at the extreme high end of the relation.

\subsubsection{Environment of HIZOA\,J0836-43}

The question that remains unanswered is: did HIZOA\,J0843-36 grow
to its current size through cannibalism in a dense environment or
accretion along filaments?  Early NIR observations found two possible
companions, further supported by the subsequent Spitzer observations.

The deep NIR IRSF survey of 2.24$\Box^\circ$ around the massive galaxy
revealed 404 galaxies, most previously unknown. Of the 44 2MASX object
in the survey area, only 16 coincided with our galaxies.  The deeper
and higher resolution IRSF images found the 28 other 2MASX objects to
either be blended (unresolved) star images or Galactic
objects. According to photometric redshifts (Jarrett et al. 2009, in prep.),
only a handful of the IRSF galaxies lie at the distance of the massive
galaxy.

The absolute magnitude distribution of the galaxies that are --
according to their photometric redshifts -- located within a volume of
radius $r = 10$\,Mpc around HIZOA\,J0836-43 are compared to four
regions of varying galaxy density at similar redshifts. This includes a
field sample, a supercluster filament, a fork apex, and a cluster
(Hercules). The comparison reveals that the population of galaxies in
the surroundings of the massive galaxy is constituted of an
overdensity of low stellar mass galaxies ($M^o_{K{_s}} \ga -23\fm0$)
with only a handful of brighter galaxies. Even fewer than in the
comparison {\sl field} sample.

Hence, the massive galaxy HIZOA\,J0836-43 lives in a low $L^*$ density
environment. That may have allowed its formation and survival,
enabling it to evolve into the unusual LIRG starburst galaxy we observe
today.  It also seems consistent with the observation that HI-massive
galaxies are preferentially found in low density regions.

Inspection of the global larger-scale structures in which HIZOA\,J0836-43 is
embedded finds this galaxy to lie in a slightly enhanced density of
fainter galaxies within a (newly identified) filamentary structure
encircling a void (Cluver 2009). The conditions appear favorable for
the galaxy to have grown this large HI disk, both through accretion of
gas due to minor merging, as well as infall of gas along the filament.

\section{Next Steps}

This is the only known HI-massive galaxy undergoing an infrared-luminous starburst. It appears to be in the process of building a disk, fueled by its large gas reservoir. We have obtained millimeter observations with MOPRA to
complement the infrared study of the galaxy's star formation phase and have detected the very broad CO (1-0) line of HIZOA\,J0836-43. We find a total molecular gas mass (H$_2$ + Helium, where the He fraction is 1.38) of 4.5$\times{10}^{9}$\MSUN\ corresponding to a H$_2$-to-HI gas-mass fraction of $\sim$6\%. We, however, require deeper observations to obtain the sensitivity to detect the low velocity CO gas and map the extent of molecular gas in the disk. This will provide clues regarding the galaxy's past and present evolution.

A future deep HI survey with MeerKAT (or ASKAP) will shed further
light on the environment of the galaxy with its association of low
stellar mass, but possibly HI-rich galaxies. Previous simulations of MeerKAT performance in the ZOA in the Great Attractor region  (Kraan-Korteweg et al. 2009) have proven its effectiveness in answering these questions on reasonable timescales. 

Such an HI-survey will furthermore give substance to the findings by
Kraan-Korteweg (in prep.) of the existence of a massive overdensity
lurking in the ZOA in Vela at $v_{hel} \sim 20000$\kms. This
overdensity is a complete surprise. Some suggestions of its existence
are seen in the reconstructed density maps based on the MASX Redshift
Survey (Erdogdu et al. 2006). If confirmed, this would have major
implications for the the dynamics in the nearby Universe. It may well
help explain the large-scale flows towards that direction found by Hudson
et al. (2009).


\bigskip

{\bf Acknowledgment:} Financial support from the National Research Foundation and the University of Cape Town is gratefully acknowledged.

\end{document}